\documentclass[letterpaper,showpacs, onecolumn,eqsecnum,superscriptaddress,floatfix]{revtex4}

\usepackage{amssymb}
\usepackage{amsmath}
\usepackage{latexsym}
\usepackage{epsfig}
\usepackage{graphicx}
\usepackage{color}
\usepackage[linktocpage]{hyperref}

\begin{document}

\title{Electromagnetic partner of the gravitational signal during accretion onto black holes}

\author{Juan Carlos Degollado} 
\email[]{jcdegollado@ciencias.unam.mx}
\affiliation{Departamento de F\'isica da Universidade de Aveiro and I3N,
Campus de Santiago, 3810-183 Aveiro, Portugal}
\affiliation{Departamento de Ciencias Computacionales,
Centro Universitario de Ciencias Exactas e Ingenier\'ia, Universidad de Guadalajara\\
Av. Revoluci\'on 1500, Colonia Ol\'impica C.P. 44430, Guadalajara, Jalisco, M\'exico}

\author{Victor Gualajara}
\email[]{vgualajara@cucea.udg.mx}
\affiliation{Centro Universitario de los Valles, Universidad de Guadalajara\\
Carretera Guadalajara-Ameca, Km. 45.5, C.P. 46600, Ameca, Jalisco, M\'exico}

\author{Claudia Moreno} 
\email[]{claudia.moreno@cucei.udg.mx} 
\affiliation{Departamento de Matem\'aticas,
Centro Universitario de Ciencias Exactas e Ingenier\'ia, Universidad de Guadalajara}

\author{Dar\'{\i}o N\'u\~nez}
\email[]{nunez@nucleares.unam.mx}
\affiliation{Instituto de Ciencias Nucleares, Universidad Nacional
  Aut\'onoma de M\'exico, Circuito Exterior C.U., A.P. 70-543,
  M\'exico D.F. 04510, M\'exico}


\date{\today}


\begin{abstract}
We investigate the generation of electromagnetic and gravitational radiation in the vicinity of 
a perturbed Schwarzschild black hole. The gravitational perturbations and the electromagnetic field are studied 
by solving the Teukolsky master equation with sources, which we take to be locally charged, radially infalling, matter. 
Our results show that, in addition to the gravitational wave generated as the matter falls into the black hole, there 
is also a burst of electromagnetic radiation. This electromagnetic field has a characteristic set of quasinormal frequencies, 
and the gravitational radiation has the quasinormal frequencies of a Schwarzschild black hole. This scenario allows us to compare 
the gravitational and electromagnetic signals that are generated by a common source.  

\end{abstract}




\maketitle


\section{Introduction}
\label{sec:introduction}

In recent years there has been a significant amount of work performed, globally, in an effort
to detect gravitational radiation. It is expected that within the next few years 
interferometric detectors will be able to detect the gravitational waves emitted from events like binary black hole, or neutron star, mergers and 
exploding, or collapsing, stars \cite{Abbott:2009ij}. 
One of the most challenging problems for the detectors, putting aside the technical obstacles,
is the extraction of the waveform from the large amount of noise generated 
in the detection process. 

Several models for electromagnetic signals correlated with the gravitational wave emissions in
astrophysical systems have been proposed \cite{Bode:2009mt}.
In this paper we describe a process in which the emission 
of gravitational waves would be accompanied by an electromagnetic signal, the infall of
charged particles onto a black hole.
Since information carried gravitationally is complementary to that carried
electromagnetically, we can learn a great deal about the generating event and its environment. For example, because the companion 
gravitational wave passes through matter almost unaffected, its properties would describe the conditions at which both signals were initially emitted.  One  
could then derive an almost redshift independent signature, see \cite{Singer:2014qca, Stubbs:2007mk} and references therein.

The program of simulating multi-messenger signals has been considered for some time in different
scenarios, and by means of several techniques. Concerning black hole spacetimes, the dynamics of
black holes immersed in a magnetic field was
considered for example in \cite{Mosta:2009rr, Palenzuela:2009yr, Palenzuela:2009hx, Moesta:2011bn}. 
In \cite{Sotani:2013iha}, a study of the coupling of axial gravitational and electromagnetic
perturbations for black holes and neutron stars was performed. It was found that for neutron stars
the ratio of the energy carried by the electromagnetic waves to the energy carried by gravitational
waves is proportional to
the square of the magnetic field of the neutron star. 
The same authors extended
their study  in \cite{Sotani:2014fia} to consider polar gravitational perturbations since these may
drive longer emission rates. Moreover, a program to study, 
numerically, the collision of charged black holes was initiated in \cite{Alcubierre:2009ij,Zilhao:2012gp}. In \cite{Zilhao:2013nda} 
it was found that there is a scaling between the gravitational signal and the electromagnetic one, where the scale factor is driven by 
the charge-mass ratio of the black holes.

Current estimates suggest that the gravitational wave signal most likely to be observed is the
\emph{chirp} from the in-spiral of two compact objects, such as neutron stars or black holes in a
close binary. It seems reasonable to expect that, after the coalescence of a compact binary, an 
accretion disk forms around the remnant black hole, and that this accretion disk persists for a long period of time, 
emitting gravitational as well as electromagnetic radiation. The latter because the disks surrounding black holes are sources for magnetic fields, 
as was described in the seminal work of Shakura and Sunyaev \cite{Shakura:1972te}. 

One can make further assumptions about the charge transport in the disk. For example one might also assume that the negative charges in the disk travel
faster than the positive charges, as discussed in \cite{deDiego:2004ar}, which would result in an effective motion
of charged matter. In our model, building on our previous results in \cite{Degollado:2009rw, 
Nunez:2010ra,Nunez:2011ej}, we will take the volume elements to be charged, and we make the
simplifying assumption that they move along geodesics; \emph{i.e} we neglect the back-reaction. As we show below, such 
matter does indeed generate a gravitational signal as well as an electromagnetic one, once it reaches the black hole horizon. 
We then attempt to determine correlations between the two signals. 
 
The paper is organized as follows: In 
section \ref{sec:foundations} we present the gravitational perturbation equations, as well as the Maxwell equations within the Newman Penrose null tetrad formalism. In 
section \ref{sec:Schwarz} we focus on the Schwarzschild background, described in 
horizon penetrating coordinates and write down the outgoing 
perturbation and electromagnetic equations explicitly.
In section \ref{sec:results} we introduce the numerical procedure to solve the equations and 
present the resulting waveforms of both signals. Finally in section\ref{sec:conclusions}, we give some concluding remarks. We adopt geometric units, $c = G = 1$, and the metric signature $(-, +, +, +)$.


\section{Foundations: Newman Penrose formalism}
\label{sec:foundations}

In \cite{Teukolsky:1973ha} Teukolsky used the spinor formulation introduced by Newman and Penrose
\cite{Newman:1962a,Newman:1966ub} to find an equation that describes neutrino, electromagnetic,
and scalar fields, as well as gravitational perturbations, on a Kerr background. The main
idea behind this formulation
lies in the choice of a basis consisting of null vectors.


A complete review of the subject can be found in \cite{Chandrasekhar83} and \cite{Degollado:2011gi}.
Teukolsky's master equation which describes a perturbed black hole, includes a term describing the source of the perturbation. 
In this work we take the source of the perturbation to be a set of infalling particles following geodesics in a non-spherical distribution.

\subsection{Gravitational perturbation}

The Einstein equations, in the Newman - Penrose formalism, consist of five 
equations for the Weyl scalars $\Psi_0, \Psi_1, \Psi_2, \Psi_3, \Psi_4$. We are interested in the scalar $\Psi_4$ that reflects the outgoing radiation. This scalar is given by
%
\begin{equation}
\Psi_4\equiv-C_{\mu\nu\lambda\tau}\,k^\mu\,{m^*}^\nu\,k^\lambda\,{m^*}^\tau\ , 
\label{eq:Psi4}
\end{equation}
where $C_{\mu\nu\lambda\tau}$ is the Weyl tensor and $k^\mu, \,m^\nu$ are two null vectors; the first
one along the null cone, pointing inwards, and the second one lying in the plane perpendicular to the light cone. 
In \cite{Teukolsky:1973ha} Teukolsky derived
a decoupled equation for the perturbation of $\Psi_4$ in terms of the spin
coefficients valid for any Type D metric. 
The equation for the perturbation $\Psi_4^{(1)}$ has the form:
\begin{eqnarray}
&&[\left({\bf \Delta} -4\,\mu_s - {\mu_s}^* - 3\,\gamma_s +{\gamma_s}^*\right)\,\left({\bf D} + 
\rho_s -4\,\epsilon_s \right) - \left({\bf \delta^*} - 3\,\alpha_s - {\beta_s}^* - 4\,\pi_s +
{\tau_s}^* \right)\,\left( {\bf \delta} -4\,\beta_s +\tau_s  \right) 
\nonumber \\
&& + 3 \Psi_2 ] \, {\Psi_4}^{(1)} = - 4\pi  \,T_4, 
\label{eq:PertPsi4}
\end{eqnarray}
where ${\bf \Delta}, {\bf D}$ and ${\bf \delta}$ are derivative operators projected along the
null directions of the tetrad components, $\mu, \gamma, \rho, \epsilon, \beta, \pi, \tau$ are the spin coefficients in the Schwarzschild metric, $*$ denotes the complex conjugate, and $\Psi_2\equiv
-C_{\mu\nu\lambda\tau}\,l^\mu\,m^\nu\,{m^*}^\lambda\,k^\tau$ is the only non zero
component of the background metric. The null vector $l^\mu$ points outwards along the light cone.
The source term $T_4$, is composed of operators acting on the projections of the stress-energy tensor
${T}_{\mu\nu}$ that triggers the gravitational response along the tetrad:
\begin{equation}
T_4 ={\cal{{\hat T}}}^{k\,k}\,{T}_{k\,k} + {\cal{{\hat T}}}^{k\,m^*}\,{T}_{k\,m^*} 
    + {\cal{{\hat T}}}^{m^*\,m^*}\,{T}_{m^*\,m^*}. \label{eq:T4g}
\end{equation}
The operators ${\cal{{\hat T}}}^{ab}$ have the explicit form
\begin{eqnarray}
{\cal{{\hat T}}}^{k\,k}&=&\left(-{\bf \delta^*} +  3\,\alpha_s + {\beta_s}^* + 4\,\pi_s
- {\tau_s}^* \right)\,\left({\bf \delta^*} -2\,\alpha_s - 2\,{\beta_s}^* +
{\tau_s}^* \right), \nonumber \\
{\cal{{\hat T}}}^{k\,m^*}&=&\left({\bf \Delta} - 4\,\mu_s - {\mu_s}^* - 3\,\gamma_s
+{\gamma_s}^* \right)\, \left({\bf \delta^*}-2\,\alpha_s +{\tau_s}^*\right)  \nonumber \\
&&+ \left({\bf \delta^*} - 3\,\alpha_s - {\beta_s}^* - 4\,\pi_s + {\tau_s}^*
\right)\,\left({\bf \Delta} -  2{\mu_s}^* - \gamma_s \right), \nonumber \\
{\cal{{\hat T}}}^{m^*\,m^*}&=&-\left({\bf \Delta} - 4\,\mu_s - {\mu_s}^* - 3\,\gamma_s
+{\gamma_s}^* \right)\,\left({\bf \Delta} -{\mu_s}^* - 2\,\gamma_s +
2\,{\gamma_s}^* \right). \label{eq:op_tau}
\end{eqnarray}
and  $T_{kk}=T_{\mu\nu}k^{\mu}k^{\nu}$, $T_{km^*}=T_{\mu\nu}k^{\mu}m^{*\nu}$, $T_{m^* m^*}=T_{\mu\nu}m^{*\mu}m^{*\nu}$ are the projections of the tensor $T_{\mu \nu}$, on the corresponding null vector.


\subsection{Maxwell equations}

The electromagnetic field, in the Newman Penrose formulation, is expressed in terms of three complex quantities 
which are projections of the Faraday tensor $F_{\mu\nu}$:
\begin{equation}
\phi_0 \equiv F_{\mu\nu} l^{\mu}m^{\nu} \ , \quad \phi_1 \equiv\frac{1}{2} F_{\mu\nu}
(l^{\mu}k^{\nu}+m^{*\mu}m^{\nu}) \ , \quad \phi_2 \equiv F_{\mu\nu} m^{*\mu}k^{\nu} \ .
\end{equation}
The scalars $\phi_0$ and $\phi_2$ represent the ingoing and outgoing electromagnetic components.
It turns out that $\phi_2$ is the electromagnetic counterpart to the outgoing gravitational radiation. Notice that,
in the case of the electromagnetic field, we are describing the actual field, which is already taken to be a perturbation 
on the background.

We then obtain a set of equations for the scalars $\phi$'s by projecting the Maxwell equations ${F^{\mu\nu}}_{;\nu}=4\,\pi\,J^\mu$ onto the null tetrad. 
Furthermore, following \cite{Teukolsky:1973ha} the equation for $\phi_2$ can be decoupled to get an equation for the outgoing component of the electromagnetic field:
\begin{eqnarray}
&&\left[\left(-\,{\bf
\Delta} + 2\,\mu_s + {\mu_s}^* + \gamma_s - {\gamma_s}^*\right)\,\left(-\,{\bf D} - \rho_s +
2\,\epsilon_s \right) -
\left(-\,{\bf \delta}^* +
\alpha_s + {\beta_s}^* + 2\,\pi_s - {\tau_s}^*\right)\,\left(-\,{\bf \delta} + 2\,\beta_s -
\tau_s \right)\right]\,\phi_2=4\,\pi\,J_2 \ ,\nonumber \\
&&
\label{eq:phi2}
\end{eqnarray}
where the source term
$J_2$ is expressed in terms of operators acting on the projections of the current vector along the null tetrad
\begin{eqnarray}
J_2&=&\left(-{\bf
\Delta} + 2\,\mu_s + {\mu_s}^* + \gamma_s - {\gamma_s}^*\right)\,J_{m^*} -
\left(-{\bf \delta}^* +
\alpha_s + {\beta_s}^* + 2\,\pi_s - {\tau_s}^*\right)\,J_k.
\end{eqnarray}
Here $J_k, J_m$ are the projections of the current vector  $J_\mu$, on the respective null vector. In this way, we arrive at two equations describing the gravitational, Eq.~(\ref{eq:PertPsi4}) and electromagnetic Eq.~(\ref{eq:phi2}) outgoing radiation, for a given source described by ${T}_{\mu\,\nu}$ and $J^\mu$.


\section{Schwarzschild background}
\label{sec:Schwarz}

We will write the gravitational perturbation equation for $\Psi_4^{(1)}$ and the Maxwell equation for $\phi_2$ in a
background given by a static black hole. We take this to be the Schwarzschild metric , written in Kerr-Schild
coordinates in order to  avoid geometric singularities at the horizon. 
\begin{equation}
ds^2=-\left(1-\frac{2M}{r}\right)dt^2 + \left(1+\frac{2M}{r}\right)dr^2 + \frac{4M}{r} drdt+r^2
(d\theta^2+ \sin^2\theta d\varphi^2) ,
\end{equation}
and define the tetrad components as
\begin{equation}
 l^{\mu}=\left(\frac{1}{2}+\frac{M}{r},\frac{1}{2}-\frac{M}{r},0,0  \right)\ , \quad k^{\mu} =
\left(1,-1,0,0 \right) \ , \quad m^{\mu}=\frac{1}{r\sqrt{2}}(0,0,1,i\csc\theta) \ .
\label{eq:tetrad}
\end{equation}
The normalization is such that $l^\mu\,k_\mu=-m^\mu\,{m^*}_\mu=-1$. The directional derivatives are    
${\bf D}=l^\mu\,\partial_\mu, {\bf \Delta}=\kappa^\mu\,\partial_\mu$ and ${\bf
\delta}=m^\mu\,\partial_\mu$. By direct substitution, we get the non-zero spin coefficients for this metric: 
\begin{eqnarray}
\mu_s=\frac{1}{r} \ , \quad \rho_s=\frac{r-2\,M}{2\,r^2} \ , \quad 
\epsilon_s=-\frac{M}{2\,r^2} \ , \quad 
\alpha_s=\frac{\cot\theta}{2\,\sqrt{2}\,r} \ , \quad \beta_s=-\alpha_s \ .
\label{eq:spincoef}
\end{eqnarray}
The non-vanishing Weyl component $\Psi_2=\frac{M}{r^3}$.

\subsection{Gravitational radiation equation}

The substitution of the directional derivatives and the spin coefficients \eqref{eq:spincoef} into \eqref{eq:PertPsi4}
yields the differential equation:
\begin{equation}
\left[{\square}_{tr}^{\Psi} + \frac{1}{r^2}\square^{-2}_{\theta\,\varphi} \right]
r\Psi_{4}^{(1)}=16\pi\,r\,{T_4} \ ,
\label{eq:pertPhi1}
\end{equation}
where the radial-temporal operator is
\begin{equation}
{\square}^{\Psi}_{tr}=-\left(1+ \frac{2M}{r}\right)\,\frac{\partial^2}{\partial t^2} + 
\left(1-\frac{2\,M}{r}\right)\,\frac{\partial^2}{\partial r^2} +
 \frac{4M}{r}\frac{\partial^2}{\partial t
\partial r} + 2\,\left(\frac{2}{r}+ \frac{M}{r^2}\right)\,\frac{\partial}{\partial t} +
2\,\left( \frac{2}{r} - \frac{M}{r^2}\right)\,\frac{\partial}{\partial r} + 
2\frac{M}{r^3}~.
\label{eq:Pert1rt}
\end{equation}
and the angular part
\begin{eqnarray}
\square^{-2}_{\theta\varphi}&=& \frac{\partial^2}{\partial \theta^2} +
\frac1{\sin^2\theta}\,\frac{\partial^2}{\partial \varphi^2} + \cot\theta\,\frac{\partial}{\partial
\theta} - 4\,i\,\frac{\cos\theta}{\sin^2\theta}\,\frac{\partial}{\partial \varphi} -
2\,\frac{1+\cos^2\theta}{\sin^2\theta}~. \label{eq:Pertang}
\end{eqnarray}
Following \cite{Degollado:2009rw}, we assume that $\Psi_{4}^{(1)}$ may be written in terms of
spherical harmonics of spin weight $s=-2$ as
\begin{equation}
\Psi_{4}^{(1)}=\sum\limits_{\ell m}\, \frac{ R^{G}_{\ell,m}(t,r)}{r}\,{Y_{-2}}^{\ell,m}(\theta,\phi)
\ ,
\label{eq:Psi_lm}
\end{equation}
where $\ell=0,1,2...$ and $|m|<\ell$.
It can be seen, for instance in \cite{Goldberg:1966uu,Chandrasekhar83}, that ${Y_{-2}}^{\ell,m}$ are
eigenfunctions of the angular operator \eqref{eq:Pertang}
\begin{equation}
\square^{-2}_{\theta\varphi}\,{Y_{-2}}^{\ell,m}=
-\left(\ell-1\right)\,\left(\ell+2\right)\,{Y_{-2}}^{\ell,m} \ .
\label{eq:Ym2}
\end{equation}
After replacing the angular operator in \eqref{eq:pertPhi1} by its eigenvalue, multiplying by the
complex conjugate ${Y^*_{-2}}^{\ell,m} $ and integrating over the sphere, Eq.~\eqref{eq:pertPhi1}
we arrive at an ordinary differential equation for each radial function
\begin{eqnarray}
&&-\left(1+\frac{2M}{r}\right)\,\frac{\partial^2 R^{G}_{\ell,m}}{\partial t^2} + 
\left(1-\frac{2M}{r}\right)\,\frac{\partial^2 R^{G}_{\ell,m}}{\partial r^2} +
\frac{4M}{r}\,\frac{\partial^2
R^{G}_{\ell,m}}{\partial t \partial r} + 2\,\left(\frac{2}{r}+\frac{M}{r^2}\right)\,\frac{\partial
R^{G}_{\ell,m}}{\partial t} + 2\,\left(\frac{2}{r}- \frac{M}{r^2}\right)\,\frac{\partial
R^{G}_{\ell,m}}{\partial r} 
\nonumber\\
&& +\left(2\frac{M}{r^3}- \frac{\left(\ell-1\right)\left(\ell+2\right)}{r^2}\right)\,R^{G}_{\ell,m}
= 16\pi \,r\,{T_{\ell,m}} . \label{eq:pert_rg}
\end{eqnarray}
The source term is $T_{\ell,m} = \int \,T_{4}\, {Y^*_{-2}}^{\ell,m} \, \sin\theta\,d\theta\,
d\varphi $. 

In order to obtain a numerical solution of this second order
equation we transform it into a first order system by defining a set auxiliary variables:
\begin{equation}
\psi^{G}_{\ell}=\partial_r\,R^{G}_{\ell} \ , \quad  \quad
\pi_{\ell}^G=\left(1+\frac{2M}r\right)\,\partial_t\,R^{G}_{\ell} -\frac{2M}r\,\psi^{G}_{\ell}\ ,
\label{eq:psig}
\end{equation}
which yields
\begin{equation}
\partial_t\,R^{G}_{\ell}=\frac{1}{r+2\,M}\left(r\,\pi^{G}_{\ell} + 2\,M\,\psi^{G}_{\ell} \right)~,
\label{eq:evolRg}
\end{equation}
\begin{eqnarray}\label{eq:evolpsig}
 \partial_t\,\psi^{G}_{\ell}&=&\partial_r\left(\frac{1}{r+2\,M}\left(r\,\pi^{G}_{\ell} +
2\,M\,\psi^{G}_{\ell}
\right)\right)\\ \nonumber
&=&\frac{1}{r+2\,M}\left(r\,\partial_r\,\pi^{G}_{\ell} + 2\,M\,\partial_r\,\psi^{G}_{\ell}
\right)
+ \frac{2\,M}{(r+2\,M)^2}\left(\pi^{G}_{\ell} - \psi^{G}_{\ell} \right) \ ,
\end{eqnarray}
\begin{eqnarray}
\partial_t\,\pi^{G}_{\ell,m}&=&\frac{1}{r+2\,M}\left(2\,M\,\partial_r\,\pi^{G}_{\ell,m} +
r\,\partial_r\,\psi^{G}_{\ell,m} \right) +
\frac{2}{r\,(r+2\,M)^2}\left(\left(2\,r^2+5\,M\,r+4\,M^2\right)\,\pi^{G}_{\ell,m} +
\left(r+4\,M\right)\,\left(2\,r+3\,M\right)\psi^{G}_{\ell,m} \right) \nonumber \\
&&+ \left(2\frac{M}{r^3} -
\frac{\left(\ell-1\right)\,\left(\ell+2\right)}{r^2}\right)\,R^{G}_{\ell,m} 
-16\pi \,r\,{T_{\ell,m}}
\ .
\label{eq:evolpig}
\end{eqnarray}

This system is solved numerically to obtain $\pi_\ell, \psi_\ell$ and $R_\ell$, as the response to $T_{\ell,m}$. In section IV we show the numerical results. 

\subsection{Electromagnetic field equation}

Substituting the directional derivatives and the spin coefficients \eqref{eq:spincoef} into
equation \eqref{eq:phi2} for $\phi_2$ we obtain

\begin{equation}
\left[{\square}_{tr}^{\phi} + \frac{1}{r^2}{}\square^{-1}_{\theta\,\varphi} \right] r\phi_{2}= -8\pi
rJ_{2},
\label{eq:pertPhi2}
\end{equation}
where
\begin{equation}
{\square}^{\phi}_{tr}=-\left(1  + \frac{2\,M}{r} 
\right)\,\frac{
\partial^2}{\partial t^2} + \left(1-\frac{2M}{r}\right)\,
\frac{\partial^2}{\partial r^2} + \frac{4\,M}{r}\,\frac{\partial^2}{\partial t \partial r} 
 +  \frac{2}{r} \,\frac{\partial}{\partial t} + 
\frac{2}{r}  \,\frac{\partial}{\partial r} \ ,
\label{eq:phi2rt}
\end{equation}
\begin{eqnarray}
\square^{-1}_{\theta\varphi}&=& \frac{\partial^2}{\partial \theta^2} +
\frac1{\sin^2\theta}\,\frac{\partial^2}{\partial \varphi^2} + \cot\theta\,\frac{\partial}{\partial
\theta} - 2\,i\,\frac{\cos\theta}{\sin^2\theta}\,\frac{\partial}{\partial \varphi} -
\frac{1}{\sin^2\theta} \ . 
\end{eqnarray}

As was done for the gravitational perturbation, we expand $\phi_2$ in spherical harmonics. However, in this case
we use a spin weight $s=-1$. The choice of the spin weight is not arbitrary; in fact it is
given by the spin of the field considered \cite{Newman:1966ub,Goldberg:1966uu}.
\begin{equation}
 \phi_2=\sum_{\ell} \frac{R^{E}_{\ell, m}(t,r)}{r} Y^{\ell, m}_{-1}(\theta,\varphi)\ .
\label{eq:phi2_lm}
\end{equation}
After replacing this expansion in \eqref{eq:pertPhi2} we get an equation for each mode
\begin{eqnarray}
&&-\left(1  + \frac{2\,M}{r} 
\right)\,\frac{\partial^2 R^{E}_{\ell, m}(t,r)}{\partial t^2} + \left(1-\frac{2M }{r}\right)
\frac{\partial^2 R^{E}_{\ell, m}(t,r)}{\partial r^2} + \frac{4\,M}{r}\,\frac{\partial^2 R^{E}_{\ell,
m}(t,r)}{\partial t \partial r} 
 +  \frac{2}{r} \,\frac{\partial R^{E}_{\ell, m}(t,r)}{\partial t}\\ \nonumber 
&& + 
\frac{2}{r}  \,\frac{\partial R^{E}_{\ell, m}(t,r)}{\partial r}
- \left(\frac{ \ell\,\left(\ell+1\right) }{r^2}\right)R^{E}_{\ell, m}(t,r)=-8\pi rJ_{\ell, m},
\end{eqnarray}
where the source term is expressed as $J_{\ell,m} = \int \,J_{2}\, {Y^*_{-1}}^{\ell,m} \,
\sin\theta\,d\theta\, d\varphi $. 

We define the corresponding first order auxiliary variables as
$\psi^{E}_{\ell}=\partial_r\,R^{E}_{\ell}$ and   
$\pi^{E}_{\ell}=\left(1+\frac{2M}r\right)\,\partial_t\,R^{E}_{\ell} -\frac{2M}r\,\psi^{E}_{\ell}$. 
%
The equations for $\partial_t R_{\ell}^{E}$ and $\partial_t\psi^{E}$ 
have the same functional form as
the gravitational case
\eqref{eq:evolRg} and \eqref{eq:evolpsig}. The equation for $\pi^{E}_{\ell}$ on the other hand  is
\begin{eqnarray}
 \partial_t\pi^{E}_{\ell} &=& \frac{1}{r+2M} \left(r\partial_r\psi^{E}_{\ell}+
2M\partial_r\pi^{E}_{\ell} 
\right)+\frac{4M^2}{r(r+2M)^2}(\pi^{E}_{\ell}-\psi^{E}_{\ell}) \nonumber \\
&&+\frac{2}{r} \left[
\frac{1}{r+2M}(r\pi^{E}_{\ell}+2M\psi^{E}_{\ell}) \right]
+\frac{2}{r}\psi^{E}_{\ell}  
-\frac{\ell (\ell+1)}{r^2}R^{E}_{\ell} +8\pi rJ_{\ell,m} \ .
\end{eqnarray}

In the next section, we give a description of the accreting matter model.
\subsection{Sources of the gravitational perturbation and the electromagnetic field}

We shall take a collection of charged particles that behave like dust to be the source of the perturbation. For simplicity each particle has the same mass $m$ and charge $e$. We also assume that the particles can not be created or destroyed. 

Let the number density in their rest frame be $n$ (the number of particles per unit of volume);
then the rest mass density is given by $\rho = m n$.
Let us further assume that the charged particles are of the same species, with a constant charge-mass ratio throughout the cloud. Then the electric current induced by the motion of the particles is 
\begin{equation}
J^{\mu}_{el}= q\rho u^{\mu} ,
\label{eq:jmu}
\end{equation}
where $q$ is the charge to mass ratio $e/m$ and $u^{\mu}$ is the four
velocity of the particles. The stress energy tensor of the system is given by
\begin{equation}
T_{\mu\nu}=\rho u_{\mu}u_{\nu} \ .
\label{eq:tmunu}
\end{equation}
We assume that the particles are falling radially into the black
hole, and we recall that in our approximation the moving particles induce an electromagnetic field, but this field does not backreact on the spacetime.
The four velocity is given by $u^{\mu}= (u^{0}(t,r),u^{1}(t,r),0,0)$.
The dynamics of the particles is governed by the conservation of the stress energy tensor and the
conservation of the four current $J^{\mu}:= \rho u^{\mu}$. 
The conservation of the mass four current, \emph{the continuity equation} $J^{\mu}{}_{;\mu}=0$
implies 
\begin{equation}
\partial_{t} (\sqrt{-g}\rho u^{0})+\partial_{r} (\sqrt{-g}\rho u^{1})=0 \ .
\label{eq:cons}
\end{equation}

The conservation of the stress energy tensor implies that the particles follow geodesics,
$u^{\nu}u^{\mu}{}_{;\nu}=0$. Using the normalization of the four velocity $u^{\mu}u_{\mu}=-1$ the
geodesics equation can be integrated to obtain the components of the four velocity,
\begin{equation}
u^1=\pm\,\sqrt{E^2 - 1 + 2\frac{M}r}~, \hspace{1cm} u^0=\frac{E\,r + 2\,M\,u^1}{r-2\,M}~,
\label{eqs:vel}
\end{equation}
where $E$ is a constant of motion related to the energy of each particle.
We choose the minus sign in the square root of $u^{1}$, indicating that the particles are in-falling
into the black hole.

It has been shown in \cite{Degollado:2009rw} that with the decomposition of the density in
spherical harmonics:
\begin{equation}
\rho=\sum\limits_{lm}\,\rho_{l,m}(t,r)\,{Y_0}^{l,m}(\theta,\phi) \ , 
\label{eq:rho}
\end{equation}
it is possible to transform \eqref{eq:cons} into a set of equations for each mode 
\begin{equation}
\partial_t\,\rho_{l,m} +
v^r\,\partial_r\,\rho_{l,m}+2\frac{E^2-1+\frac{3\,M}{2\,r}}{r\,\left(E^2-1+\frac{2\,M}{r}\right)}\,
v^r\,\rho_{l,m}=0~, \label{eq:evolrho}
\end{equation}
where we used the 3 velocity defined as $v^{r} = u^{1}/u^{0}$. Notice that the configuration \eqref{eq:rho} is not necessarily spherically symmetric.
  
Let us focus on the electromagnetic source for $\phi_2$. Given the current vector \eqref{eq:jmu}, and
the expansion \eqref{eq:rho} the only non-vanishing projection is along the vector $k^{\mu}$ 
(Notice that $k_\mu\,u^\mu=-\left(u^0+u^1\right)$)
\begin{eqnarray}
 J_2 &=& \frac{1}{r\sqrt{2}}(\partial_{\theta}-i\,
\csc\theta\partial_{\varphi})J_{k} \nonumber\\
&=&  
\frac{q}{r\sqrt{2}}  ( u^{0}+u^{1}) \sum\limits_{\ell m}\,\rho_{\ell, m}  {\bar\eth}_{0}
Y_{0}^{\ell, m}(\theta,\varphi) \\\nonumber
& =& 
-\frac{q}{r\sqrt{2}}  ( u^{0}+u^{1})\sqrt{\ell(\ell+1)} \sum\limits_{\ell, m}\,\rho_{\ell, m} 
Y_{-1}^{\ell, m}(\theta,\varphi) \ ,
\end{eqnarray}
where we have used the fact that the operator {\it eth-bar}, ${\bar \eth}$, lowers
the spin weight of the spherical harmonics in the form:
\begin{equation}
{\bar \eth}_s\,{Y_s}^{l,m}=-\sqrt{\left(l+s\right)\,\left(l-s+1\right)}\,{Y_{s-1}}^{\ell,m} \ .
\label{eq:ethl_Y}
\end{equation}

Then, we integrate over the sphere to get
\begin{equation}
J_{\ell, m} = -\frac{q}{r\sqrt{2}}  ( u^{0}+u^{1})\sqrt{\ell(\ell+1)}\,\rho_{\ell,m}\ . \label{eq:J}
\end{equation}
From here it is straightforward to see that the monopole does not excite the electromagnetic field. Notice also that it is enough to solve the radial-temporal components to get the waveform.

In a similar way, 
using these expressions for the density and the four velocity in the stress energy tensor, Eq.~(\ref{eq:tmunu}),
we compute the needed projections on the tetrad and act with the operators ${\cal{{\hat T}}}^{ab}$, from Eqs.~(\ref{eq:op_tau}).
The details of this derivation were presented in \cite{{Degollado:2009rw}}. The final result is that
the source term for the gravitational perturbation takes the form
\begin{equation}
T_4=
-\frac{(u^0+u^1)^2}{2\,r^2}\sum\limits_{\ell, m}\,\rho_{\ell,m}(t,r)\,\sqrt{\left(\ell-1\right)\,\ell\,
\left(\ell+1\right)\,\left(\ell+2\right)}\,{Y_{-2}}^{\ell,m} \ .
\label{eq:T4}
\end{equation}
and consequently, after an integration over the solid angle, and taking into account the
orthonormality of the spherical harmonics, we obtain the following expression for the 
gravitational source:
\begin{equation}
 T_{\ell, m} =
-\frac{(u^0+u^1)^2}{2\,r^2}\,\rho_{\ell,m}(t,r)\,\sqrt{\left(\ell-1\right)\,\ell\,
\left(\ell+1\right)\,\left(\ell+2\right)} \ .
\label{eq:T_lm}
\end{equation}

Any initial matter distribution can be expanded in terms of spherical harmonics: According to \eqref{eq:T_lm} and \eqref{eq:J} the monopole mode will not generate any gravitational or electromagnetic reaction; the dipole mode will generate an electromagnetic but not a gravitational response, and the quadrupole and higher modes generate both type of responses.

 Furthermore, each mode of matter excites only the 
corresponding electromagnetic and gravitational mode with same harmonic number $l$. The 
decomposition \eqref{eq:T_lm} and \eqref{eq:J} allows us to study, with a 1 dimensional numerical code, any
radially infalling charged dust matter distribution and its gravitational and electromagnetic reaction. 
Notice also that there is a degeneracy with respect to the $m$ modes. This degeneracy comes from the
fact that the background is spherically symmetric and that the fluid movement is restricted to be radial.

\subsection{Time evolution}

We solve the equations for the gravitational perturbation with sources plus the equations for the electromagnetic field by using
the Method Of Lines (MoL). The detailed description
of the code is given in \cite{Degollado:2009rw}. Here we just summarize the main aspects. 
The numerical code evolves the first order variables with a third order Runge Kutta
integrator with a fourth order spatial stencil in a domain $r\in [r_{min},r_{max}]$. 
Since we are using Kerr-Schild type coordinates $r_{min}$ lies inside the event horizon. Typically we choose
$r_{min}=1.5M$ and $r_{max}=2000M$. As usual, we introduce a small sixth order dissipation to get rid off high frequency modes. At the boundaries we impose the condition that all the incoming waves (as given by the characteristic fields) vanish. 

In all the simulations presented we use as initial data a gaussian packet in the density. This represents a non spherical shell of particles falling into the hole. 
We solve the equation for the rest mass density and look for the gravitational and electromagnetic response. 
The gravitational and electromagnetic waveforms are then extracted at a fixed $r=r_o$ radius.


\section{Results and discussions}
\label{sec:results}

In the following we shall report the behaviour of the gravitational signal and its electromagnetic
counterpart when charged dust particles fall into the black hole. 

We have considered for simplicity, a shell of matter described by a single 
spherical harmonic mode
\begin{equation}
 \rho(t,r,\theta,\varphi)= \rho_{\ell,m}(t,r) Y_{0}^{\ell,m}(\theta,\varphi) .
\end{equation}
We will take the $\ell=1,2$ modes since these modes yields the main contribution to
the quadrupole gravitational radiation and the dipolar electromagnetic radiation which are expected
to be dominant. The modes with $\ell > 2$  will be relevant for higher gravitational and
electromagnetic multipoles. 

We have used as initial data for the radial distribution a Gaussian centred at $r_{cg}$: 
\begin{equation}
 \rho_{\ell,m}(0,r)=\rho_0 e^{-(r-r_{cg})^2/2\sigma^2 } \ , 
 \label{eq:igauss}
\end{equation}
with $\rho_0=5\times10^{-3}$, $r_{cg}=100M$ and $\sigma=0.5M$, but we will vary this width in
some experiments as described below. The gravitational and
electromagnetic functions, $R^G, R^E$, are set to zero, as well as their time derivatives.
Our results indicate that the effect of this choice on the gravitational and electromagnetic waveforms has
negligible. Finally, the outer boundary was set far enough from the horizon to ensure that any possible incoming radiation has no effect on our results.

\subsection{Waveforms}

Our numerical results confirm the expected proportionality of the wavefunction with the charge. 
We found that the electromagnetic radial wave function is homogeneous of degree 1 in the charge i.e.
$R^{E}(t,r_o;\lambda q)=\lambda  R^{E}(t,r_o;q)$, 
indeed we found that $R^{E}$ is linear in the charge $q$.

\begin{widetext}
\begin{figure}[!ht]
\begin{center}
 \includegraphics[width=0.35\textwidth,clip,angle=-90]{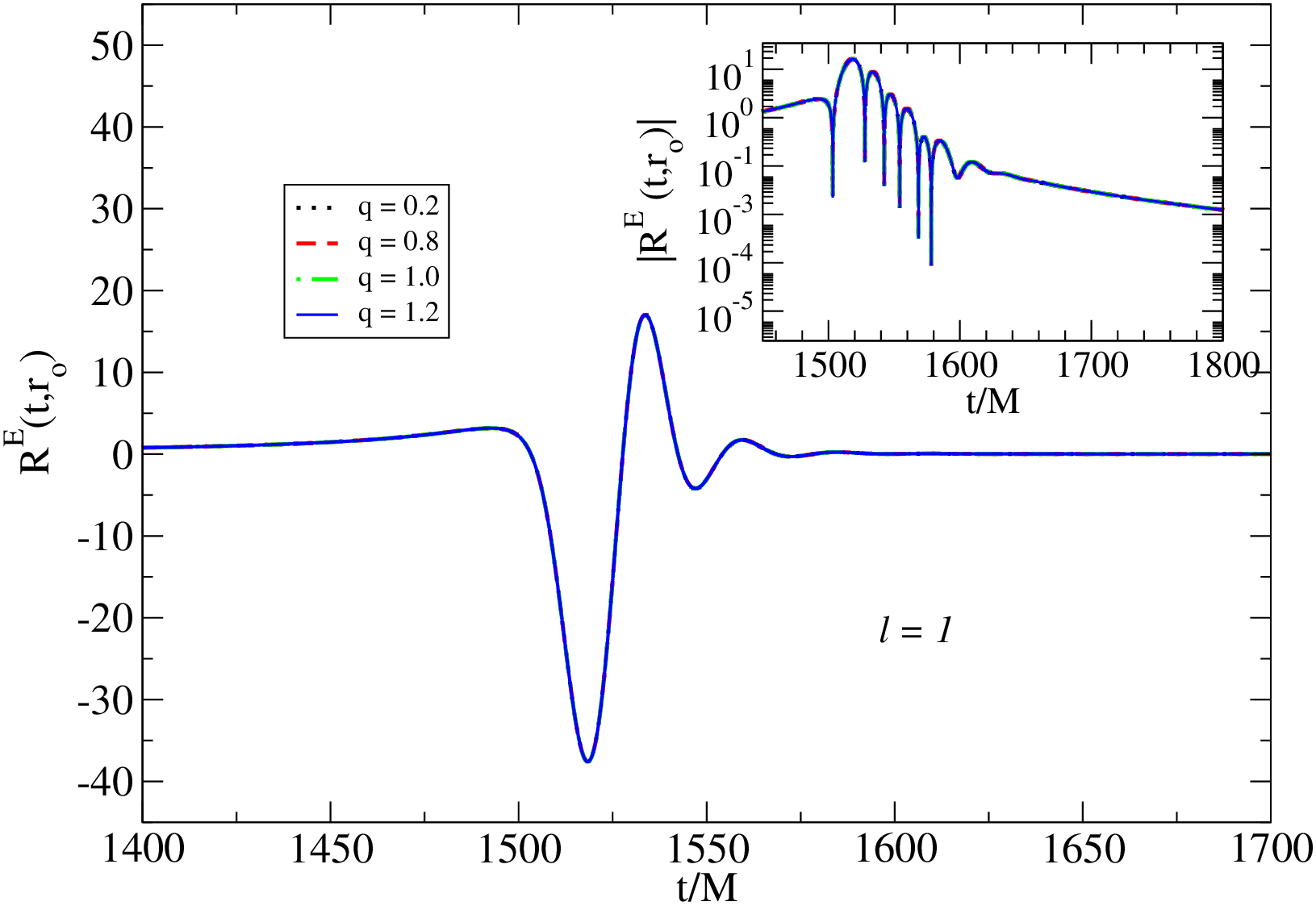}
\includegraphics[width=0.35\textwidth,clip,angle=-90]{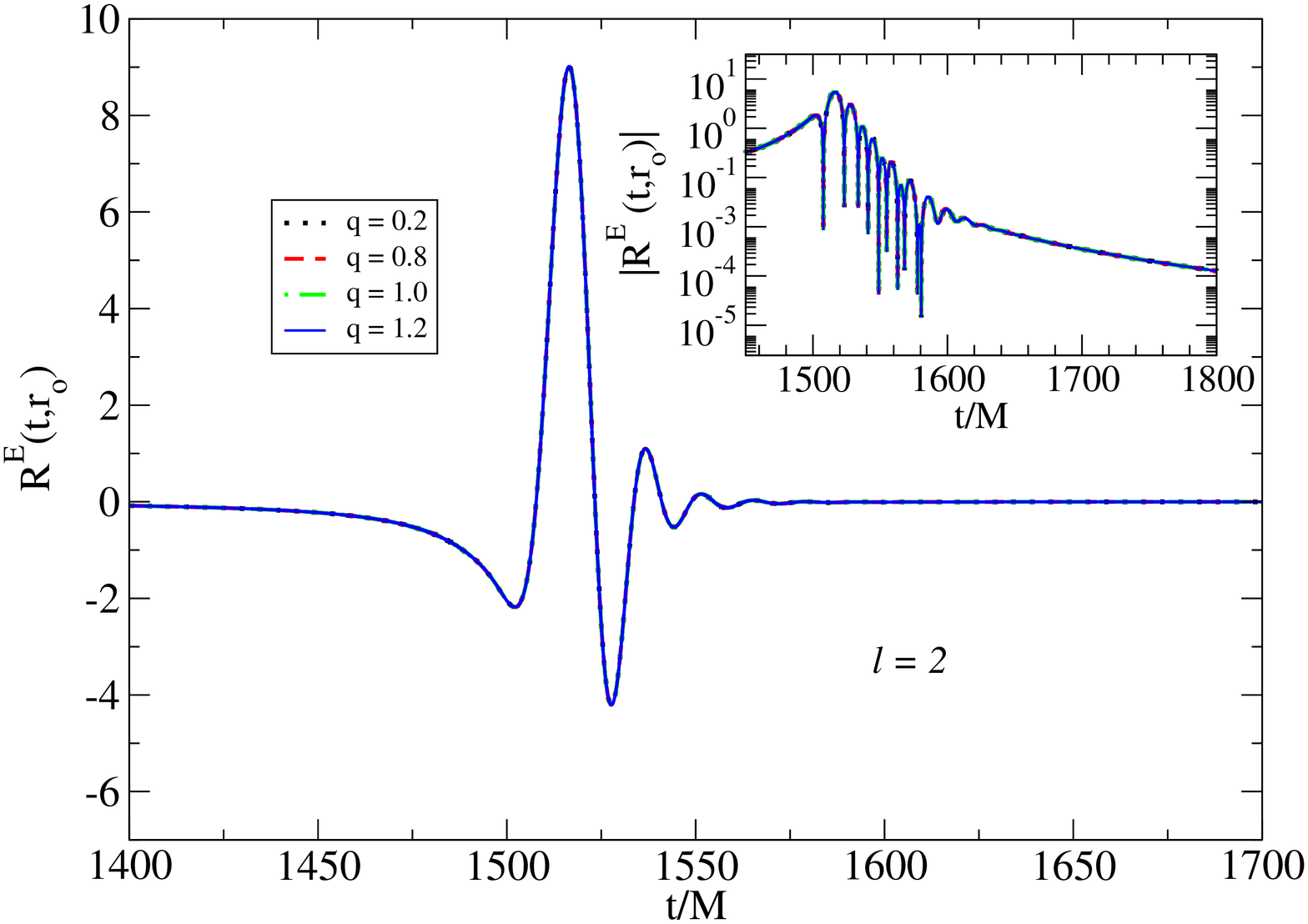}
\caption{Electromagnetic signal for $\ell=1$ and $\ell=2$ modes. The waveforms have been rescaled
to show the property $R^{E}(t,r_o;\lambda q)=\lambda  R^{E}(t,r_o;q)$. 
}
\label{fig:obsR1_elec_qs} 
\end{center}
\end{figure}
\end{widetext}

In Fig.~\ref{fig:obsR1_elec_qs} we display the radial waveforms as obtained at 
 $r_o=1000M$. 
In the left panel we display the dipolar component $\ell=1$, and in the right panel, the quadrupolar $\ell=2$. 
In the inset we plot the absolute value on a logarithmic scale to show the different
stages of the signal: the initial burst due to the initial data, the quasinormal ringing
and the tail. In these figures the waveforms have been rescaled to show the dependence of $R^{E}$ on $q$.

While the ring down frequency is determined by the quasinormal ringing, it is interesting to note
that the amplitude can be derived from a scaling law. This behaviour might be related
with another scaling observed in \cite{Zilhao:2013nda}. In that work, the authors considered the collision
of two charged black holes with opposite charge $Q$ and mass $M$ such that the resultant black hole was
of the Schwarzschild type. It was found that the emitted electromagnetic waveform in the final black
hole scales as $Q(1+Q^2/M^2)^{(1/2)}$. For a distant observer, however, the scenario is that there is
an electromagnetic signal coming from an almost stationary black hole, very similar to our final set-up when the charged particles have fallen into the hole.

\begin{figure}[!ht]
\begin{center}
\includegraphics[width=0.5\textwidth,clip,angle=-90]{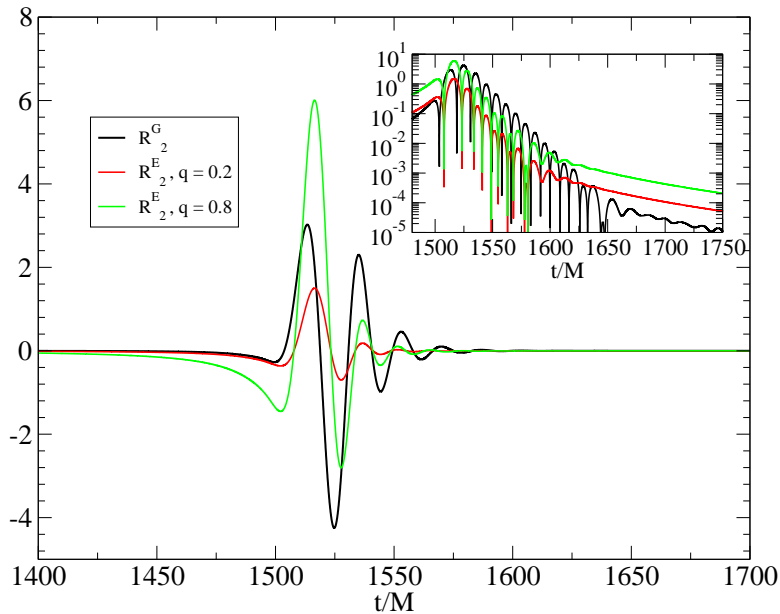}
\caption{Radial profiles of the gravitational ($R^G_{\ell}$) and electromagnetic  ($R^{E}_{\ell}$) signals for the 
quadrupolar $\ell=2$ mode. The extraction radius is $r_o=1000M$. The inset panel shows the absolute 
value in a logarithmic scale to properly distinguish the different phases; the ringdown and the tail decay. 
}
\label{fig:compEleGrav} 
\end{center}
\end{figure}

A comparison between the electromagnetic and gravitational waveforms is shown in
Fig.~\ref{fig:compEleGrav}. The figure shows the quadrupolar modes for the gravitational signal and two representative cases for the electromagnetic signal with charge $q=0.2, 0.8$.  
We find no sign of mixing between gravitational and electromagnetic frequencies. Each signal displays its
characteristic ring down frequency. The inset shows the absolute value on a logarithmic scale. 
We found that the electromagnetic waves have a specific frequency  corresponding to the $\ell=1$
quasinormal mode of electromagnetic waves in a Schwarzschild background
\cite{Cunningham78,Berti:2009kk,Konoplya:2011qq}. The frequency of the gravitational waves on the
other hand is the one associated to the quadrupolar $\ell=2$ quasinormal mode \cite{Chandrasekhar75, Price72,
Leaver:1985ax,Kokkotas:1999bd}. 
In order to find the frequencies we fit the data with a sinusoidal waveform. 
The numerical values of the corresponding frequencies are listed in table \ref{tab:qnmfreq}. The values we find coincide with the values previously reported in 
\cite{Cunningham78}. 
While it is known that quasinormal mode frequencies are in general complex, we were only interested on the oscillatory behaviour of the signal and not in their time decay. For black holes of a few solar masses $10M_{\odot}<M<10^{3}M_{\odot}$ the electromagnetic frequencies
are in the interval $8$Hz $\sim$ $800$Hz whereas the gravitational waves produced for such a range of masses are in the $12$Hz to $1.2$kHz window. As 
has been pointed out in several works, quasinormal ringing can be used to determine the intrinsic 
properties of the black hole \cite{Berti:2009kk}.
Electromagnetic waves with such low frequencies however could easily be absorbed by the interstellar
medium during its propagation and it will be almost impossible to directly detect them. 
One possible way to observe the electromagnetic waves is through the detection of its indirect effects on
the medium, such as synchrotron radiation. In this scenario, however, more information is needed about the medium. 
\begin{center}
\begin{table}
  \begin{tabular}{ |c ||c | c |} 
    \hline
 $\ell$   & $M\omega^{GW} $ & $M\omega^{EM} $\\ 
\hline 
$1$ & - & 0.248 \\ 
\hline
$2$ & 0.373 & 0.457 \\
\hline
$3$ & 0.598 & 0.655 \\
\hline
  \end{tabular}
\caption{Frequencies of the gravitational and electromagnetic waveforms obtained with a sinusoidal fit. The values match, up to the decimal figures shown, the quasi normal frequencies reported in \cite{Cunningham78}.
}

\label{tab:qnmfreq}
\end{table}
\end{center}
We studied the effect of the variation of the Gaussian half width in the gravitational and 
electromagnetic signals.
The response of an isolated black hole to the width of Gaussian initial perturbations is well known. When 
the Gaussian is very broad, no quasinormal ringing could be seen in the scattered radiation. When 
the Gaussian is made thinner ringing occurs \cite{Vishveshwara:1970zz,Kokkotas:1999bd}. Similar 
results were reported when the perturbation was caused by a Gaussian distribution of 
dust \cite{Papadopoulos:1998nc, Sotani:2005be, Nagar:2006eu, Degollado:2009rw}. 
Our results for charged dust particles point in the same direction for both 
electromagnetic and gravitational signals. 

\begin{widetext}
\begin{figure}[!ht]
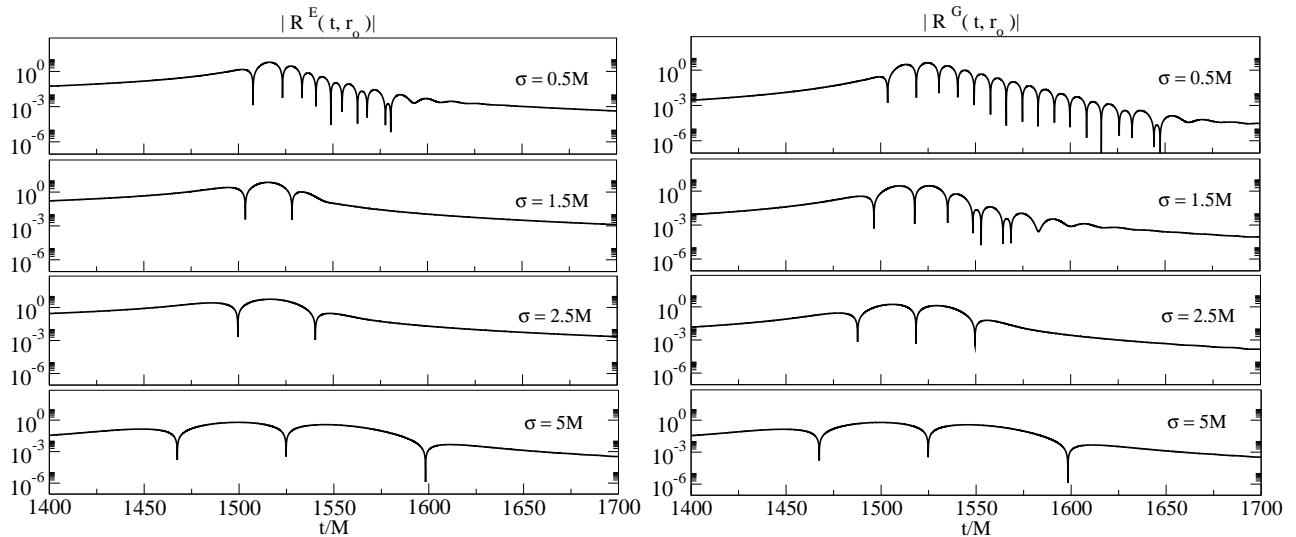

\begin{center}
 \includegraphics[width=0.47\textwidth,clip]{obsR1elecsigma.eps}
\includegraphics[width=0.47\textwidth,clip]{obsR1gravsigma.eps}
\caption{Absolute value of the electromagnetic (left) and gravitational (right) quadrupolar $\ell=2$ waveforms for different values of the Gaussian width $\sigma$. The observer is at $r_o =1000M$ and the Gaussian was centred at $r_{cg}=100M$. The charge mass ratio used in these examples was $q=0.8$. As $\sigma$ increases the lost of the ring down behaviour becomes evident for both signals.}
\label{fig:obs1_vs_sigma} 
\end{center}
\end{figure}
\end{widetext}

The waveforms for several values of the initial Gaussian width are plotted in figure \ref{fig:obs1_vs_sigma}. The left panel corresponds to the electromagnetic wave and the right to the gravitational one. The width of the Gaussian increases from top to bottom. From the figure it is clear that no ringing at all is seen for very broad Gaussians.    
%

\subsection{Energy radiated}

In terms of the multipolar decomposition \eqref{eq:Psi_lm}, the radiated gravitational energy flux 
is given by \cite{Teukolsky:1973ha}:
\begin{equation}
 P_{GW} = \frac{d}{dt}E_{GW}= \lim_{r\rightarrow\infty}\frac{1}{16\pi} \sum_{\ell,m}
|\int_{-\infty}^{t} dt' R^{G}_{\ell}(t')|^2 \ .
\label{eq:flux_gw}
\end{equation}
The energy flux carried off by outgoing electromagnetic waves in terms of the
multipolar decomposition \eqref{eq:phi2_lm} is given by
\begin{equation}
 P_{EM} = \frac{d}{dt}E_{EM}= \lim_{r\rightarrow\infty}\frac{1}{4\pi} \sum_{\ell,m}  |R^{E}_{\ell
}(t)|^2 \ .
\label{eq:flux_em}
\end{equation}

\begin{figure}[!ht]
\begin{center}
\includegraphics[width=0.5\textwidth,clip]{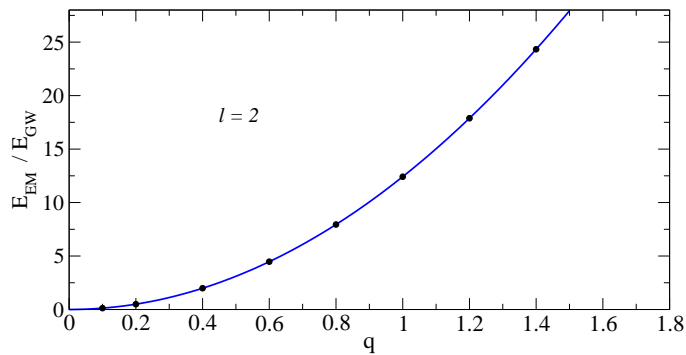}
\caption{The ratio of the electromagnetic and gravitational radiated energy 
computed from the integration of the fluxes ~\eqref{eq:flux_gw}, \eqref{eq:flux_em} (dots). The continuous 
line is a quadratic fit $E_{EM}/E_{GW} = 12.417 q^2$.}
\label{fig:E_EM_vs_q} 
\end{center}
\end{figure}

We compute the total radiated energies by direct integration of Eqs.~\eqref{eq:flux_gw}, \eqref{eq:flux_em}. 
To calculate
the total energy, we start the integration of the fluxes after some time in order 
to get rid of the radiation resulting from the initial data. 
The choice of the start time depends on the
extraction radius, typically we take it to be $t_s\sim 1400M$ for an extraction radius of $r_o=1000M$. The value of the energy does not change when we put the observers far away, indicating the validity of the far region approximation. 
In order to show the effect of the charge of the particles on both the gravitational and electromagnetic signals,
we consider the quotient between the computed energies. In Fig.~\ref{fig:E_EM_vs_q} we plot the
ratio $E^{EM}/E_{GW}$ versus 
the density of charge $q$ for $\ell=2$. We choose as initial data a gaussian pulse with $\sigma
=0.5M$ and $r_{cg}=100M$. As expected, when the value of the charge increases the amount of
electromagnetic energy also increases. For 
small values of the charge the gravitational energy dominates but as the charge grows, it is the electromagnetic energy 
which dominates. We found that there is a relation between the two energies and fit their quotient
with a quadratic 
function of the form $E_{EM}/E_{GW} = a\, q^2$. We found that the numerical value of $a$ depends on
the initial width of the gaussian. In Ref.~\cite{Degollado:2009rw} it was shown that the energy of
the gravitational wave varies monotonically with the width of the gaussian. This behaviour is 
related with the possibility of the initial data to induce quasinormal modes. Our result indicates
that the electromagnetic energy varies in such a way that the quotient $E_{EM}/E_{GW}$
depends quadratically on $q$.
We also have found that the value of $a$, for a fixed width that produces quasinormal ringing,
depends on the angular momentum number $\ell$ with $a=12.417, 11.128, 10.928$ for $\ell=2,3,4$
respectively.

\section{Conclusions}
\label{sec:conclusions}

In this paper we studied the combined, electromagnetic and gravitational, response due to a shell 
of charged particles falling radially, along geodesics, into a Schwarzschild black hole. Varying the angular distribution of the particles allowed us 
to produce changes in the multipolar components of the infalling matter, and to excite gravitational and electromagnetic emission. 

In this way, we found that the falling matter triggers two signals, one electromagnetic and one gravitational. 
The gravitational signal displays the characteristic behaviour of a perturbed black hole; an
oscillating 
phase after an initial burst and a final power law decay. The electromagnetic waves show the same 
qualitative behaviour. However, we found no direct coupling between the frequencies of both types
of 
waves. The gravitational signal corresponds to the quasinormal ringing and the electromagnetic 
corresponds to a electromagnetic field scattered from a Schwarzschild black hole.
The absence of any type of coupling between the frequencies may be due to the fact that we neglected the 
interaction between the charge of the particles and the electromagnetic field in the background. We expect that the coupling will be manifest in a model which takes into account the backreaction of the spacetime.

We found that if the initial distribution of dust is very broad no quasinormal ringing appears. 
We studied the effect of the charge of the particles on the signals emitted and found that there 
is a linear dependence of the electromagnetic waveforms on the charge-mass ratio, $q$.
We also found that the gravitational and electromagnetic energies are related by the square of $q$. 
Within our simplified model, 
if one were able to determine the electromagnetic signal one would then be able to 
infer properties of the gravitational wave, such as the frequency and the energy content once 
the charge was determined by other means. In order to achieve this goal more involved 
models of matter must be taken into account. 

There are several directions in which the present study could 
be continued. It may be worthwhile to extend the analysis to include more realistic 
models for accretion disks, to consider not only vacuum spacetimes and to study the system presented
in this manuscript in a rotating black hole scenario. Work is this direction is under way.


\section*{Acknowledgements}

We acknowledge Jai Grover, Carlos Herdeiro and Carlos Palenzuela for a 
careful reading and useful comments to improve the manuscript.
This work was supported in part by DGAPA-UNAM grants IN115311 and IN103514 and PROSNI.
JCD acknowledges support from FCT via project No. PTDC/FIS/116625/2010. This work was also supported
by the NRHEPÐ 295189
FP7-PEOPLE-2011-IRSES grant. CM acknowledges support from FOMIX-UDG project No. 2010-10-149481.


\bibliography{referencias}

\begin{thebibliography}{35}
\expandafter\ifx\csname natexlab\endcsname\relax\def\natexlab#1{#1}\fi
\expandafter\ifx\csname bibnamefont\endcsname\relax
  \def\bibnamefont#1{#1}\fi
\expandafter\ifx\csname bibfnamefont\endcsname\relax
  \def\bibfnamefont#1{#1}\fi
\expandafter\ifx\csname citenamefont\endcsname\relax
  \def\citenamefont#1{#1}\fi
\expandafter\ifx\csname url\endcsname\relax
  \def\url#1{\texttt{#1}}\fi
\expandafter\ifx\csname urlprefix\endcsname\relax\def\urlprefix{URL }\fi
\providecommand{\bibinfo}[2]{#2}
\providecommand{\eprint}[2][]{\url{#2}}

\bibitem[{\citenamefont{et~al}(2009)}]{Abbott:2009ij}
\bibinfo{author}{\bibfnamefont{B.~P.~A.} \bibnamefont{et~al}},
  \bibinfo{journal}{Rep. Prog. Phys.} \textbf{\bibinfo{volume}{72}},
  \bibinfo{pages}{076901} (\bibinfo{year}{2009}), \eprint{arXiv:0711.3041}.

\bibitem[{\citenamefont{Bode et~al.}(2010)\citenamefont{Bode, Haas, Bogdanovic,
  Laguna, and Shoemaker}}]{Bode:2009mt}
\bibinfo{author}{\bibfnamefont{T.}~\bibnamefont{Bode}},
  \bibinfo{author}{\bibfnamefont{R.}~\bibnamefont{Haas}},
  \bibinfo{author}{\bibfnamefont{T.}~\bibnamefont{Bogdanovic}},
  \bibinfo{author}{\bibfnamefont{P.}~\bibnamefont{Laguna}}, \bibnamefont{and}
  \bibinfo{author}{\bibfnamefont{D.}~\bibnamefont{Shoemaker}},
  \bibinfo{journal}{Astrophys.J.} \textbf{\bibinfo{volume}{715}},
  \bibinfo{pages}{1117} (\bibinfo{year}{2010}), \eprint{0912.0087}.

\bibitem[{\citenamefont{Singer et~al.}(2014)\citenamefont{Singer, Price, Farr,
  Urban, Pankow et~al.}}]{Singer:2014qca}
\bibinfo{author}{\bibfnamefont{L.~P.} \bibnamefont{Singer}},
  \bibinfo{author}{\bibfnamefont{L.~R.} \bibnamefont{Price}},
  \bibinfo{author}{\bibfnamefont{B.}~\bibnamefont{Farr}},
  \bibinfo{author}{\bibfnamefont{A.~L.} \bibnamefont{Urban}},
  \bibinfo{author}{\bibfnamefont{C.}~\bibnamefont{Pankow}},
  \bibnamefont{et~al.} (\bibinfo{year}{2014}), \eprint{1404.5623}.

\bibitem[{\citenamefont{Stubbs}(2008)}]{Stubbs:2007mk}
\bibinfo{author}{\bibfnamefont{C.~W.} \bibnamefont{Stubbs}},
  \bibinfo{journal}{Class.Quant.Grav.} \textbf{\bibinfo{volume}{25}},
  \bibinfo{pages}{184033} (\bibinfo{year}{2008}), \eprint{0712.2598}.

\bibitem[{\citenamefont{Mosta et~al.}(2010)\citenamefont{Mosta, Palenzuela,
  Rezzolla, Lehner, Yoshida et~al.}}]{Mosta:2009rr}
\bibinfo{author}{\bibfnamefont{P.}~\bibnamefont{Mosta}},
  \bibinfo{author}{\bibfnamefont{C.}~\bibnamefont{Palenzuela}},
  \bibinfo{author}{\bibfnamefont{L.}~\bibnamefont{Rezzolla}},
  \bibinfo{author}{\bibfnamefont{L.}~\bibnamefont{Lehner}},
  \bibinfo{author}{\bibfnamefont{S.}~\bibnamefont{Yoshida}},
  \bibnamefont{et~al.}, \bibinfo{journal}{Phys.Rev.}
  \textbf{\bibinfo{volume}{D81}}, \bibinfo{pages}{064017}
  (\bibinfo{year}{2010}), \eprint{0912.2330}.

\bibitem[{\citenamefont{Palenzuela et~al.}(2009)\citenamefont{Palenzuela,
  Anderson, Lehner, Liebling, and Neilsen}}]{Palenzuela:2009yr}
\bibinfo{author}{\bibfnamefont{C.}~\bibnamefont{Palenzuela}},
  \bibinfo{author}{\bibfnamefont{M.}~\bibnamefont{Anderson}},
  \bibinfo{author}{\bibfnamefont{L.}~\bibnamefont{Lehner}},
  \bibinfo{author}{\bibfnamefont{S.~L.} \bibnamefont{Liebling}},
  \bibnamefont{and} \bibinfo{author}{\bibfnamefont{D.}~\bibnamefont{Neilsen}}
  (\bibinfo{year}{2009}), \eprint{astro-ph/0905.1121}.

\bibitem[{\citenamefont{Palenzuela et~al.}(2010)\citenamefont{Palenzuela,
  Lehner, and Yoshida}}]{Palenzuela:2009hx}
\bibinfo{author}{\bibfnamefont{C.}~\bibnamefont{Palenzuela}},
  \bibinfo{author}{\bibfnamefont{L.}~\bibnamefont{Lehner}}, \bibnamefont{and}
  \bibinfo{author}{\bibfnamefont{S.}~\bibnamefont{Yoshida}},
  \bibinfo{journal}{Phys.Rev.} \textbf{\bibinfo{volume}{D81}},
  \bibinfo{pages}{084007} (\bibinfo{year}{2010}), \eprint{0911.3889}.

\bibitem[{\citenamefont{Moesta et~al.}(2012)\citenamefont{Moesta, Alic,
  Rezzolla, Zanotti, and Palenzuela}}]{Moesta:2011bn}
\bibinfo{author}{\bibfnamefont{P.}~\bibnamefont{Moesta}},
  \bibinfo{author}{\bibfnamefont{D.}~\bibnamefont{Alic}},
  \bibinfo{author}{\bibfnamefont{L.}~\bibnamefont{Rezzolla}},
  \bibinfo{author}{\bibfnamefont{O.}~\bibnamefont{Zanotti}}, \bibnamefont{and}
  \bibinfo{author}{\bibfnamefont{C.}~\bibnamefont{Palenzuela}},
  \bibinfo{journal}{Astrophys.J.} \textbf{\bibinfo{volume}{749}},
  \bibinfo{pages}{L32} (\bibinfo{year}{2012}), \eprint{1109.1177}.

\bibitem[{\citenamefont{Sotani et~al.}(2013)\citenamefont{Sotani, Kokkotas,
  Laguna, and Sopuerta}}]{Sotani:2013iha}
\bibinfo{author}{\bibfnamefont{H.}~\bibnamefont{Sotani}},
  \bibinfo{author}{\bibfnamefont{K.~D.} \bibnamefont{Kokkotas}},
  \bibinfo{author}{\bibfnamefont{P.}~\bibnamefont{Laguna}}, \bibnamefont{and}
  \bibinfo{author}{\bibfnamefont{C.~F.} \bibnamefont{Sopuerta}},
  \bibinfo{journal}{Phys.Rev.} \textbf{\bibinfo{volume}{D87}},
  \bibinfo{pages}{084018} (\bibinfo{year}{2013}), \eprint{1303.5641}.

\bibitem[{\citenamefont{Sotani et~al.}(2014)\citenamefont{Sotani, Kokkotas,
  Laguna, and Sopuerta}}]{Sotani:2014fia}
\bibinfo{author}{\bibfnamefont{H.}~\bibnamefont{Sotani}},
  \bibinfo{author}{\bibfnamefont{K.~D.} \bibnamefont{Kokkotas}},
  \bibinfo{author}{\bibfnamefont{P.}~\bibnamefont{Laguna}}, \bibnamefont{and}
  \bibinfo{author}{\bibfnamefont{C.~F.} \bibnamefont{Sopuerta}},
  \bibinfo{journal}{Gen.Rel.Grav.} \textbf{\bibinfo{volume}{46}},
  \bibinfo{pages}{1675} (\bibinfo{year}{2014}), \eprint{1402.0251}.

\bibitem[{\citenamefont{Alcubierre et~al.}(2009)\citenamefont{Alcubierre,
  Degollado, and Salgado}}]{Alcubierre:2009ij}
\bibinfo{author}{\bibfnamefont{M.}~\bibnamefont{Alcubierre}},
  \bibinfo{author}{\bibfnamefont{J.~C.} \bibnamefont{Degollado}},
  \bibnamefont{and} \bibinfo{author}{\bibfnamefont{M.}~\bibnamefont{Salgado}},
  \bibinfo{journal}{Phys.Rev.} \textbf{\bibinfo{volume}{D80}},
  \bibinfo{pages}{104022} (\bibinfo{year}{2009}), \eprint{0907.1151}.

\bibitem[{\citenamefont{Zilhao et~al.}(2012)\citenamefont{Zilhao, Cardoso,
  Herdeiro, Lehner, and Sperhake}}]{Zilhao:2012gp}
\bibinfo{author}{\bibfnamefont{M.}~\bibnamefont{Zilhao}},
  \bibinfo{author}{\bibfnamefont{V.}~\bibnamefont{Cardoso}},
  \bibinfo{author}{\bibfnamefont{C.}~\bibnamefont{Herdeiro}},
  \bibinfo{author}{\bibfnamefont{L.}~\bibnamefont{Lehner}}, \bibnamefont{and}
  \bibinfo{author}{\bibfnamefont{U.}~\bibnamefont{Sperhake}},
  \bibinfo{journal}{Phys.Rev.} \textbf{\bibinfo{volume}{D85}},
  \bibinfo{pages}{124062} (\bibinfo{year}{2012}), \eprint{1205.1063}.

\bibitem[{\citenamefont{Zilhao et~al.}(2013)\citenamefont{Zilhao, Cardoso,
  Herdeiro, Lehner, and Sperhake}}]{Zilhao:2013nda}
\bibinfo{author}{\bibfnamefont{M.}~\bibnamefont{Zilhao}},
  \bibinfo{author}{\bibfnamefont{V.}~\bibnamefont{Cardoso}},
  \bibinfo{author}{\bibfnamefont{C.}~\bibnamefont{Herdeiro}},
  \bibinfo{author}{\bibfnamefont{L.}~\bibnamefont{Lehner}}, \bibnamefont{and}
  \bibinfo{author}{\bibfnamefont{U.}~\bibnamefont{Sperhake}}
  (\bibinfo{year}{2013}), \eprint{1311.6483}.

\bibitem[{\citenamefont{Shakura and Sunyaev}(1973)}]{Shakura:1972te}
\bibinfo{author}{\bibfnamefont{N.}~\bibnamefont{Shakura}} \bibnamefont{and}
  \bibinfo{author}{\bibfnamefont{R.}~\bibnamefont{Sunyaev}},
  \bibinfo{journal}{Astron.Astrophys.} \textbf{\bibinfo{volume}{24}},
  \bibinfo{pages}{337} (\bibinfo{year}{1973}).

\bibitem[{\citenamefont{de~Diego et~al.}(2004)\citenamefont{de~Diego,
  Dultzin-Hacyan, Trejo, and Nunez}}]{deDiego:2004ar}
\bibinfo{author}{\bibfnamefont{J.~A.} \bibnamefont{de~Diego}},
  \bibinfo{author}{\bibfnamefont{D.}~\bibnamefont{Dultzin-Hacyan}},
  \bibinfo{author}{\bibfnamefont{J.~G.} \bibnamefont{Trejo}}, \bibnamefont{and}
  \bibinfo{author}{\bibfnamefont{D.}~\bibnamefont{Nunez}}
  (\bibinfo{year}{2004}), \eprint{astro-ph/0405237}.

\bibitem[{\citenamefont{Degollado et~al.}(2010)\citenamefont{Degollado, Nunez,
  and Palenzuela}}]{Degollado:2009rw}
\bibinfo{author}{\bibfnamefont{J.~C.} \bibnamefont{Degollado}},
  \bibinfo{author}{\bibfnamefont{D.}~\bibnamefont{Nunez}}, \bibnamefont{and}
  \bibinfo{author}{\bibfnamefont{C.}~\bibnamefont{Palenzuela}},
  \bibinfo{journal}{Gen. Rel. Grav.} \textbf{\bibinfo{volume}{42}},
  \bibinfo{pages}{1287} (\bibinfo{year}{2010}), \eprint{0903.2073}.

\bibitem[{\citenamefont{Nunez et~al.}(2010)\citenamefont{Nunez, Degollado, and
  Palenzuela}}]{Nunez:2010ra}
\bibinfo{author}{\bibfnamefont{D.}~\bibnamefont{Nunez}},
  \bibinfo{author}{\bibfnamefont{J.~C.} \bibnamefont{Degollado}},
  \bibnamefont{and}
  \bibinfo{author}{\bibfnamefont{C.}~\bibnamefont{Palenzuela}},
  \bibinfo{journal}{Phys.Rev.} \textbf{\bibinfo{volume}{D81}},
  \bibinfo{pages}{064011} (\bibinfo{year}{2010}), \eprint{1002.2227}.

\bibitem[{\citenamefont{Nunez et~al.}(2011)\citenamefont{Nunez, Degollado, and
  Moreno}}]{Nunez:2011ej}
\bibinfo{author}{\bibfnamefont{D.}~\bibnamefont{Nunez}},
  \bibinfo{author}{\bibfnamefont{J.~C.} \bibnamefont{Degollado}},
  \bibnamefont{and} \bibinfo{author}{\bibfnamefont{C.}~\bibnamefont{Moreno}},
  \bibinfo{journal}{Phys.Rev.} \textbf{\bibinfo{volume}{D84}},
  \bibinfo{pages}{024043} (\bibinfo{year}{2011}), \eprint{1107.4316}.

\bibitem[{\citenamefont{Teukolsky}(1973)}]{Teukolsky:1973ha}
\bibinfo{author}{\bibfnamefont{S.~A.} \bibnamefont{Teukolsky}},
  \bibinfo{journal}{Astrophys.J.} \textbf{\bibinfo{volume}{185}},
  \bibinfo{pages}{635} (\bibinfo{year}{1973}).

\bibitem[{\citenamefont{Newman and Penrose}(1962)}]{Newman:1962a}
\bibinfo{author}{\bibfnamefont{E.~T.} \bibnamefont{Newman}} \bibnamefont{and}
  \bibinfo{author}{\bibfnamefont{R.}~\bibnamefont{Penrose}},
  \bibinfo{journal}{J. Math. Phys.} \textbf{\bibinfo{volume}{3}},
  \bibinfo{pages}{566} (\bibinfo{year}{1962}), \bibinfo{note}{erratum in J.
  Math. Phys. 4, 998 (1963)}.

\bibitem[{\citenamefont{Newman and Penrose}(1966)}]{Newman:1966ub}
\bibinfo{author}{\bibfnamefont{E.~T.} \bibnamefont{Newman}} \bibnamefont{and}
  \bibinfo{author}{\bibfnamefont{R.}~\bibnamefont{Penrose}},
  \bibinfo{journal}{J. Math. Phys.} \textbf{\bibinfo{volume}{7}},
  \bibinfo{pages}{863} (\bibinfo{year}{1966}).

\bibitem[{\citenamefont{Chandrasekhar}(1983)}]{Chandrasekhar83}
\bibinfo{author}{\bibfnamefont{S.}~\bibnamefont{Chandrasekhar}},
  \emph{\bibinfo{title}{The Mathematical Theory of Black Holes}}
  (\bibinfo{publisher}{Oxford University Press}, \bibinfo{address}{Oxford,
  England}, \bibinfo{year}{1983}).

\bibitem[{\citenamefont{Degollado and Nunez}(2011)}]{Degollado:2011gi}
\bibinfo{author}{\bibfnamefont{J.~C.} \bibnamefont{Degollado}}
  \bibnamefont{and} \bibinfo{author}{\bibfnamefont{D.}~\bibnamefont{Nunez}},
  \bibinfo{journal}{AIP Conf.Proc.} \textbf{\bibinfo{volume}{1473}},
  \bibinfo{pages}{3} (\bibinfo{year}{2011}).

\bibitem[{\citenamefont{Goldberg et~al.}(1967)\citenamefont{Goldberg,
  MacFarlane, Newman, Rohrlich, and Sudarshan}}]{Goldberg:1966uu}
\bibinfo{author}{\bibfnamefont{J.}~\bibnamefont{Goldberg}},
  \bibinfo{author}{\bibfnamefont{A.}~\bibnamefont{MacFarlane}},
  \bibinfo{author}{\bibfnamefont{E.}~\bibnamefont{Newman}},
  \bibinfo{author}{\bibfnamefont{F.}~\bibnamefont{Rohrlich}}, \bibnamefont{and}
  \bibinfo{author}{\bibfnamefont{E.}~\bibnamefont{Sudarshan}},
  \bibinfo{journal}{J.Math.Phys.} \textbf{\bibinfo{volume}{8}},
  \bibinfo{pages}{2155} (\bibinfo{year}{1967}).

\bibitem[{\citenamefont{{Cunningham} et~al.}(1978)\citenamefont{{Cunningham},
  {Price}, and {Moncrief}}}]{Cunningham78}
\bibinfo{author}{\bibfnamefont{C.~T.} \bibnamefont{{Cunningham}}},
  \bibinfo{author}{\bibfnamefont{R.~H.} \bibnamefont{{Price}}},
  \bibnamefont{and}
  \bibinfo{author}{\bibfnamefont{V.}~\bibnamefont{{Moncrief}}},
  \bibinfo{journal}{\apj} \textbf{\bibinfo{volume}{224}}, \bibinfo{pages}{643}
  (\bibinfo{year}{1978}).

\bibitem[{\citenamefont{Berti et~al.}(2009)\citenamefont{Berti, Cardoso, and
  Starinets}}]{Berti:2009kk}
\bibinfo{author}{\bibfnamefont{E.}~\bibnamefont{Berti}},
  \bibinfo{author}{\bibfnamefont{V.}~\bibnamefont{Cardoso}}, \bibnamefont{and}
  \bibinfo{author}{\bibfnamefont{A.~O.} \bibnamefont{Starinets}},
  \bibinfo{journal}{Class.Quant.Grav.} \textbf{\bibinfo{volume}{26}},
  \bibinfo{pages}{163001} (\bibinfo{year}{2009}), \eprint{0905.2975}.

\bibitem[{\citenamefont{Konoplya and Zhidenko}(2011)}]{Konoplya:2011qq}
\bibinfo{author}{\bibfnamefont{R.}~\bibnamefont{Konoplya}} \bibnamefont{and}
  \bibinfo{author}{\bibfnamefont{A.}~\bibnamefont{Zhidenko}},
  \bibinfo{journal}{Rev.Mod.Phys.} \textbf{\bibinfo{volume}{83}},
  \bibinfo{pages}{793} (\bibinfo{year}{2011}), \eprint{1102.4014}.

\bibitem[{\citenamefont{{Chandrasekhar} and
  {Detweiler}}(1975)}]{Chandrasekhar75}
\bibinfo{author}{\bibfnamefont{S.}~\bibnamefont{{Chandrasekhar}}}
  \bibnamefont{and}
  \bibinfo{author}{\bibfnamefont{S.}~\bibnamefont{{Detweiler}}},
  \bibinfo{journal}{Royal Society of London Proceedings Series A}
  \textbf{\bibinfo{volume}{344}}, \bibinfo{pages}{441} (\bibinfo{year}{1975}).

\bibitem[{\citenamefont{{Price}}(1972)}]{Price72}
\bibinfo{author}{\bibfnamefont{R.~H.} \bibnamefont{{Price}}},
  \bibinfo{journal}{\prd} \textbf{\bibinfo{volume}{5}}, \bibinfo{pages}{2419}
  (\bibinfo{year}{1972}).

\bibitem[{\citenamefont{Leaver}(1985)}]{Leaver:1985ax}
\bibinfo{author}{\bibfnamefont{E.}~\bibnamefont{Leaver}},
  \bibinfo{journal}{Proc.Roy.Soc.Lond.} \textbf{\bibinfo{volume}{A402}},
  \bibinfo{pages}{285} (\bibinfo{year}{1985}).

\bibitem[{\citenamefont{Kokkotas and Schmidt}(1999)}]{Kokkotas:1999bd}
\bibinfo{author}{\bibfnamefont{K.~D.} \bibnamefont{Kokkotas}} \bibnamefont{and}
  \bibinfo{author}{\bibfnamefont{B.~G.} \bibnamefont{Schmidt}},
  \bibinfo{journal}{Living Rev.Rel.} \textbf{\bibinfo{volume}{2}},
  \bibinfo{pages}{2} (\bibinfo{year}{1999}), \eprint{gr-qc/9909058}.

\bibitem[{\citenamefont{Vishveshwara}(1970)}]{Vishveshwara:1970zz}
\bibinfo{author}{\bibfnamefont{C.}~\bibnamefont{Vishveshwara}},
  \bibinfo{journal}{Nature} \textbf{\bibinfo{volume}{227}},
  \bibinfo{pages}{936} (\bibinfo{year}{1970}).

\bibitem[{\citenamefont{Papadopoulos and Font}(1999)}]{Papadopoulos:1998nc}
\bibinfo{author}{\bibfnamefont{P.}~\bibnamefont{Papadopoulos}}
  \bibnamefont{and} \bibinfo{author}{\bibfnamefont{J.~A.} \bibnamefont{Font}},
  \bibinfo{journal}{Phys.Rev.} \textbf{\bibinfo{volume}{D59}},
  \bibinfo{pages}{044014} (\bibinfo{year}{1999}), \eprint{gr-qc/9808054}.

\bibitem[{\citenamefont{Sotani and Saijo}(2006)}]{Sotani:2005be}
\bibinfo{author}{\bibfnamefont{H.}~\bibnamefont{Sotani}} \bibnamefont{and}
  \bibinfo{author}{\bibfnamefont{M.}~\bibnamefont{Saijo}},
  \bibinfo{journal}{Phys.Rev.} \textbf{\bibinfo{volume}{D74}},
  \bibinfo{pages}{024001} (\bibinfo{year}{2006}), \eprint{gr-qc/0507030}.

\bibitem[{\citenamefont{Nagar et~al.}(2007)\citenamefont{Nagar, Zanotti, Font,
  and Rezzolla}}]{Nagar:2006eu}
\bibinfo{author}{\bibfnamefont{A.}~\bibnamefont{Nagar}},
  \bibinfo{author}{\bibfnamefont{O.}~\bibnamefont{Zanotti}},
  \bibinfo{author}{\bibfnamefont{J.~A.} \bibnamefont{Font}}, \bibnamefont{and}
  \bibinfo{author}{\bibfnamefont{L.}~\bibnamefont{Rezzolla}},
  \bibinfo{journal}{Phys.Rev.} \textbf{\bibinfo{volume}{D75}},
  \bibinfo{pages}{044016} (\bibinfo{year}{2007}), \eprint{gr-qc/0610131}.

\end{thebibliography}

 
\end{document}